\documentclass[10pt,a4paper]{article}
\usepackage{amsmath,amssymb,latexsym}

\def\AFOUR{%
\setlength{\textheight}{8.5in}%
\setlength{\textwidth}{5.75in}%
\setlength{\topmargin}{-0.375in}%
\hoffset=-.5in%
\renewcommand{\baselinestretch}{1.17}%
\setlength{\parskip}{6pt plus 2pt}%
}

\AFOUR                                           

\expandafter\ifx\csname amssym.def\endcsname\relax \else\endinput\fi
\expandafter\edef\csname amssym.def\endcsname{%
       \catcode`\noexpand\@=\the\catcode`\@\space}
\catcode`\@=11
\def\undefine#1{\let#1\undefined}
\def\newsymbol#1#2#3#4#5{\let\next@\relax
 \ifnum#2=\@ne\let\next@\msafam@\else
 \ifnum#2=\tw@\let\next@\msbfam@\fi\fi
 \mathchardef#1="#3\next@#4#5}
\def\mathhexbox@#1#2#3{\relax
 \ifmmode\mathpalette{}{\m@th\mathchar"#1#2#3}%
 \else\leavevmode\hbox{$\m@th\mathchar"#1#2#3$}\fi}
\def\hexnumber@#1{\ifcase#1 0\or 1\or 2\or 3\or 4\or 5\or 6\or 7\or 8\or
 9\or A\or B\or C\or D\or E\or F\fi}

\font\tenmsa=msam10
\font\sevenmsa=msam7
\font\fivemsa=msam5
\newfam\msafam
\textfont\msafam=\tenmsa
\scriptfont\msafam=\sevenmsa
\scriptscriptfont\msafam=\fivemsa
\edef\msafam@{\hexnumber@\msafam}
\mathchardef\dabar@"0\msafam@39
\def\dashrightarrow{\mathrel{\dabar@\dabar@\mathchar"0\msafam@4B}}
\def\dashleftarrow{\mathrel{\mathchar"0\msafam@4C\dabar@\dabar@}}

\def\ulcorner{\delimiter"4\msafam@70\msafam@70 }
\def\urcorner{\delimiter"5\msafam@71\msafam@71 }
\def\llcorner{\delimiter"4\msafam@78\msafam@78 }
\def\lrcorner{\delimiter"5\msafam@79\msafam@79 }
\def\yen{{\mathhexbox@\msafam@55}}
\def\checkmark{{\mathhexbox@\msafam@58}}
\def\circledR{{\mathhexbox@\msafam@72}}
\def\maltese{{\mathhexbox@\msafam@7A}}
\def\circledS{{\mathhexbox@\msafam@73}}

\font\tenmsb=msbm10
\font\sevenmsb=msbm7
\font\fivemsb=msbm5
\newfam\msbfam
\textfont\msbfam=\tenmsb
\scriptfont\msbfam=\sevenmsb
\scriptscriptfont\msbfam=\fivemsb
\edef\msbfam@{\hexnumber@\msbfam}
\def\Bbb#1{{\fam\msbfam\relax#1}}
\def\widehat#1{\setbox\z@\hbox{$\m@th#1$}%
 \ifdim\wd\z@>\tw@ em\mathaccent"0\msbfam@5B{#1}%
 \else\mathaccent"0362{#1}\fi}
\def\widetilde#1{\setbox\z@\hbox{$\m@th#1$}%
 \ifdim\wd\z@>\tw@ em\mathaccent"0\msbfam@5D{#1}%
 \else\mathaccent"0365{#1}\fi}
\font\teneufm=eufm10
\font\seveneufm=eufm7
\font\fiveeufm=eufm5
\newfam\eufmfam
\textfont\eufmfam=\teneufm
\scriptfont\eufmfam=\seveneufm
\scriptscriptfont\eufmfam=\fiveeufm
\def\frak#1{{\fam\eufmfam\relax#1}}

\csname amssym.def\endcsname

\makeatletter
\def\section{\@startsection {section}{1}{\z@}{-3.5ex plus -1ex minus
 -.2ex}{2.3ex plus .2ex}{\large\sc}}
\def\subsection{\@startsection{subsection}{2}{\z@}{-3.25ex plus -1ex minus
 -.2ex}{1.5ex plus .2ex}{\normalsize\sc}}
\makeatother

\makeatletter
\@addtoreset{equation}{section}

\makeatother

\newcommand{\nc}{\newcommand}
\newcommand{\rnc}{\renewcommand}

\nc{\chap}[1]{{\clearpage}%
\begin{center}%
{\noindent\underline{\large\sc #1}}{\addcontentsline{toc}{section}{#1}}%
\end{center}%
{\vspace*{0.3cm}}}

\nc{\be}{\begin{equation}}
\nc{\ee}{\end{equation}}
\nc{\bea}{\begin{eqnarray}}
\nc{\eea}{\end{eqnarray}}

\nc{\trac}[2]{{\textstyle\frac{#1}{#2}}}

\nc{\ex}[1]{\mbox{e}^{\,\textstyle#1}}

\nc{\CC}{\Bbb{C}}
\nc{\HH}{\Bbb{H}}
\nc{\PP}{\Bbb{P}}
\nc{\RR}{\Bbb{R}}
\nc{\ZZ}{\Bbb{Z}}
\nc{\II}{\Bbb{I}}
\nc{\EE}{\Bbb{E}}
\nc{\TT}{\Bbb{T}}
\nc{\DD}{\mathrm{I}\!\mathrm{D}}

\rnc{\a}{\alpha}
\rnc{\b}{\beta}
\rnc{\d}{\delta}
\nc{\ga}{\gamma}
\nc{\la}{\lambda}
\nc{\f}{\phi}
\nc{\e}{\eta}
\rnc{\c}{\chi}
\nc{\eps}{\epsilon}
\nc{\om}{\omega}
\nc{\Om}{\Omega}

\nc{\symx}{\circledS}
\newsymbol\smallsmile 1360
\newsymbol\smallfrown 1361
\nc{\ad}{\mathop{\mbox{ad}}\nolimits}
\nc{\tr}{\mathop{\mbox{tr}}\nolimits}
\nc{\Tr}{\mathop{\mbox{Tr}}\nolimits}
\nc{\Det}{\mathop{\mbox{Det}}\nolimits}
\rnc{\det}{\mathop{\mbox{det}}\nolimits}
\nc{\rk}{\mathop{\mbox{rk}}\nolimits}
\nc{\del}{\partial}
\nc{\diag}{\mathop{\mbox{diag}}\nolimits}
\nc{\ra}{\rightarrow}
\nc{\Ra}{\Rightarrow}
\nc{\LRa}{\Leftrightarrow}
\nc{\lra}{\leftrightarrow}
\nc{\ot}{\otimes}
\rnc{\ss}{\subset}
\nc{\nul}{\noindent\underline}
\nc{\non}{\nonumber\\}
\nc{\mat}[4]{\left(\begin{array}{cc}#1&#2\\#3&#4\end{array}\right)}
\rnc{\lg}{\frak{g}}
\nc{\G}[3]{\Gamma^{#1}_{\;{#2}{#3}}}
\nc{\nam}{\nabla_{\mu}}
\nc{\nan}{\nabla_{\nu}}
\nc{\dx}{\dot{x}}
\nc{\dxl}{\dot{x}^{\la}}
\nc{\dxm}{\dot{x}^{\mu}}
\nc{\dxn}{\dot{x}^{\nu}}
\nc{\ddx}{\ddot{x}}
\nc{\ddxm}{\ddot{x}^{\mu}}
\nc{\ddxn}{\ddot{x}^{\nu}}
\nc{\dxi}{\dot{\xi}}
\nc{\ddxi}{\ddot{\xi}}

\newcommand{\Exp}{\mathrm{Exp}}
\newcommand{\p}{\partial}

\providecommand{\tr}{\text{Tr}}

\providecommand{\comment}[1]{}
\begin{document}


\begin{center}
{\Large\sc Penrose Limits vs String Expansions}
\end{center}
\vspace{0.2cm}

\begin{center}
{\large\sc Matthias Blau} and {\large\sc Sebastian Weiss}
\end{center}

\vskip 0.2 cm
\centerline{\it Institut de Physique, Universit\'e de Neuch\^atel}
\centerline{\it Rue Breguet 1, CH-2000 Neuch\^atel, Switzerland} 

\vspace{1cm}

We analyse the relation between two a priori quite different expansions
of the string equations of motion and constraints in a general curved
background, namely one based on the covariant Penrose-Fermi expansion of
the metric $G_{\mu\nu}$ around a Penrose limit plane wave associated to a
null geodesic $\gamma$, and the other on the Riemann coordinate expansion
in the exact metric $G_{\mu\nu}$ of the string embedding variables
around the null geodesic $\gamma$. Starting with the observation that
there is a formal analogy between the exact string equations in a plane
wave and the first order string equations in a general background,
we show that this analogy becomes exact provided that one chooses the
background string configuration to be the null geodesic $\gamma$ itself.
We then explore the higher-order correspondence between these two
expansions and find that for a general curved background they agree
to all orders provided that one works in Fermi coordinates and in the
lightcone gauge. Requiring moreover the conformal gauge restricts one
to the usual class of (Brinkmann) backgrounds admitting simultaneously
the lightcone and the conformal gauge, without further restrictions.

\section{Introduction}

After the initial developments \cite{bfhp1,rrm,bfhp2} related to
the discovery of the maximally supersymmetric IIB plane wave and its
connection with the Penrose limit \cite{penrose,gueven}, much effort
has, in the wake of the seminal BMN paper \cite{bmn}, understandably
gone into exploring the consequences of these ideas in the context of
the AdS/CFT correspondence, eventually leading to deep new insights
into the integrable structures underlying the theories on both sides of
the correspondence. Some of these developments are described e.g.\ in
\cite{sheikh,beisert,tseytlin1}.

Along a different line, in a series of papers
\cite{bfp,mmhom,mmga,bbop1,bbop2,bfwf} we have explored
various aspects of the geometry and physics of plane waves and Penrose
limits \textit{per se}, also with the expectation that these results will
eventually lead to further insights into the gauge theory -- geometry
correspondence.  In particular, in \cite{bbop1,bbop2} we provided a
geometrically transparent and covariant characterisation of the Penrose limit
map
$(G_{\mu\nu},\gamma) \mapsto A_{ab}$ that associates to a space-time
metric $G_{\mu\nu}$ and a null geodesic $\gamma$ the wave profile $A_{ab}$
characterising the Penrose limit plane wave metric $ds^2 = 2 dx^+dx^- +
A_{ab}(x^+)x^a x^b (dx^+)^2 + \delta_{ab}\, dx^a dx^b$.  Namely, the $A_{ab}(x^+)=
-R_{+a+b}(x^+)$, which are the only non-vanishing coordinate components
of the curvature tensor of the plane wave, are at the same time
simply certain frame components of the curvature tensor of the original
metric $G_{\mu\nu}$, restricted to the null geodesic $\gamma$ (with affine
parameter $x^+$) along which the Penrose limit is taken.

In \cite{bfwf}, we used Fermi coordinates based on the null geodesic
$\gamma$ to generalise the above result to an all order covariant
expansion of a metric around its Penrose limit (covariant in the
sense that all the higher order terms are also expressed in terms
of the Riemann tensor of the original metric and its derivatives).
In the following we will refer to this expansion as the
\textit{Penrose-Fermi expansion} of a metric.

Within this clear geometric setting it is now possible to address
questions regarding the relation between the dynamics of various
objects in the original metric and its Penrose limit. In particular,
the above geometric interpretation of the Penrose limit can be
re-interpreted as providing an answer to the \begin{list}{}{\topsep=0cm}
\item[\sc Question:] What is the interpretation of the geodesic
equation in the Penrose limit plane wave (associated to the metric
$G_{\mu\nu}$ and a null geodesic $\gamma$) in terms of the original
data $(G_{\mu\nu},\gamma)$?  \item[\sc Answer:] It is simply the
transverse geodesic deviation equation for $(G_{\mu\nu},\gamma)$.
\end{list} It is then natural to next ask the same question for
strings rather than for particles.\footnote{In a similar spirit,
in \cite{bfws} we showed that scalar field probes of space-time
singularities exhibit a universal behaviour that is strictly analogous
to that of massless particle probes (i.e.\ the Penrose limit)
uncovered in \cite{bbop1,bbop2}.} \begin{list}{}{\topsep=0cm}
\item[\sc Question:] What is the interpretation of the string
equations of motion in the Penrose limit plane wave in terms of the
string equations of motion in the original metric $G_{\mu\nu}$?
\end{list} Thinking about this, one quickly realises that this will
have to be related to a (first order) expansion of the string
embedding variables $X^{\mu}(\tau,\sigma)$ around the null geodesic
$\gamma(\tau)$, the latter regarded as a string background solution
of the equations of motion and constraints in the original space-time
with metric $G_{\mu\nu}$.

Thus, in more general terms what this amounts to is a comparison of
two apparently quite different expansions of the string equations in a
curved background, an expansion of the metric itself (the Penrose-Fermi
expansion of $G_{\mu\nu}$) 
on the one hand, and an expansion of the string embedding
variables around a background string configuration (but in the exact
metric $G_{\mu\nu}$) on the other.

In order to be able to assess what the advantages (or perhaps drawbacks) are
of choosing a null geodesic as a (somewhat degenerate) string background 
configuration, we have found it useful to begin the discussion with an
analysis of the expansion of the string equations around a non-degenerate
string background configuration $X_B^\mu(\tau,\sigma)$. This is, of course,
largely classical material, the Riemann coordinate expansion of the
two-dimensional sigma-model having been discussed at length e.g.\ in 
\cite{AlvarezGaume:1981hn}, and we briefly recall this (and adapt it to the
present setting) in section 2 and appendix \ref{riem1}.

The principal difference to the discussion of \cite{AlvarezGaume:1981hn}
is that, in addition to the sigma-model equations of motion we also have
to deal with the string constraints. Then the main observation of this
section is that, to first order in an expansion around a classical string
configuration $X_B(\tau,\sigma)$ in an arbitrary curved background,
these constraints allow one to explicitly solve for the tangential
fluctuations and to completely eliminate them from the equations of
motion for the true dynamical transverse degrees of freedom. While the
result as such may not be surprising (it is essentially a consequence
of worldsheet diffeomorphism invariance), our presentation is aimed at
highlighting the analogies and differences with strings in the conformal
and lightcone gauge in plane wave backgrounds.

We pursue this analogy in section 3, where we observe first of all
that the main difference between the first order and plane wave
equations of motion for the true dynamical transverse degrees of
freedom is due to the extrinsic curvature of the background string
$X_B$. We then argue that this difference disappears, and that the
analogy becomes perfect, when one chooses the background string
configuration to be a null geodesic, $X_B(\tau,\sigma) \ra \gamma(\tau)$. 
The result of this section can
then be summarised as the answer to the question posed above.
\begin{list}{}{\topsep=0cm}
\item[\textsc{Answer:}]
The exact transverse string equations in the first order Penrose-Fermi
expansion of the metric $G_{\mu\nu}$ around $\gamma$, i.e.\ in the 
Penrose limit plane
wave metric associated to $G_{\mu\nu}$ and $\gamma$, are equivalent to
the transverse first-order string expansion equations around a 
null geodesic $\gamma$ in the original background $G_{\mu\nu}$.
\end{list}

Finally, in section 4 we address the 
\begin{list}{}{\topsep=0cm}
\item[\sc Question:] 
What can one say about the correspondence between the string
expansion on the one hand and the Penrose-Fermi expansion on the other,
established to first order in section 3, at higher orders?
\end{list}
This boils down to a comparison of two different prescriptions for how
to describe the locus of nearby strings in terms of geodesic distance
(namely via Riemann or Fermi coordinates). We show that demanding all
order equivalence of the two expansions is tantamount to the requirement
that the string be comoving with the null geodesic, and these geometric
considerations then lead to the
\begin{list}{}{\topsep=0cm}
\item[\sc Answer:]
Provided that one works in Fermi coordinates and in
the lightcone gauge, these two expansions agree to all orders.
\end{list}

This combined lightcone (worldsheet) and Fermi (space-time) gauge (i.e.\
writing the metric in Fermi coordinates) is, a priori,
always available.  Frequently, however, the lightcone gauge is imposed
in conjunction with the conformal gauge, and this requires a metric that
has a parallel null vector, as well as a coordinate system in which this
is a coordinate vector $\del_v$ \cite{hs}. We show (appendix B) that
for all such metrics the latter requirement is actually compatible with
the Fermi gauge. Since for this class of metrics canonical quantisation
becomes particularly tractable in the lightcone and conformal gauge,
this makes this all order equivalence especially appealing.

These results provide us with what seems to be a satisfactory overall
geometric picture of the relation between string dynamics in a general
curved background and in the Penrose-Fermi expansion of that background
around its Penrose limit plane wave metric.

We should also note here in passing that the idea of basing
a string expansion on an expansion around a geodesic is as such of
course not new. Such an expansion was e.g.\ considered (to first order)
in \cite{sanchezvegabh}, primarily for specific examples of
metric backgrounds, and using (for reasons we do not fully comprehend)
timelike instead of null geodesics. An expansion based on null geodesics
was considered in \cite{zheltukhin}, in the context of tensionless
strings. While formally similar, our treatment of this expansion is quite
different, both technically (using in an essential way the manifestly
covariant Riemann and Fermi coordinate expansions) and in spirit. E.g.\
we argued in \cite{bfhp2,bfp} that the Penrose limit is most naturally
understood as a particular large tension $\alpha^\prime\ra 0$ limit, and 
in the present context the Riemann coordinate (derivative) expansion we
employ can, as usual, be translated into an $\alpha^\prime$ expansion.

\section{Covariant string expansion around a regular string background
solution\label{2}}

Our point of departure is the Polyakov action 
\begin{equation}
S[X,h]=\frac{1}{2\pi \alpha^\prime} \int d^2\sigma 
\sqrt{h} h^{ij} G_{\mu\nu}(X) \p_i X^\mu \p_j X^\nu, 
\end{equation}
for a string moving in the $D$-dimensional
curved space-time background described by the metric $G_{\mu\nu}$, with
$X^{\mu}=X^{\mu}(\tau,\sigma)$ the string
embedding
variables corresponding to the target space coordinates $x^{\mu}$, and
$h_{ij}$ the worldsheet metric. Throughout this paper, with the exception of
section 4, we work in the conformal gauge $h_{ij}=e^{\phi}\eta_{ij}$, 
leading to the sigma-model action (the conformal factor $e^{\phi}$ drops out
of all subsequent equations)
\begin{equation}\label{confaction}
S[X]=\frac{1}{2\pi \alpha^\prime} \int  
d^2\sigma G_{\mu\nu}(X) \p^i X^\mu \p_i X^\nu.
\end{equation}
The equations of motion (e.o.m.)
\begin{equation}\label{euler}
\nabla^i\p_i X^\mu=\p^i\p_i X^\mu +\Gamma_{\nu\lambda}^\mu(X)
\p^i X^\nu\p_i X^\lambda=0
\end{equation}
have to be supplemented by the constraints 
\begin{equation}\label{constr}
G_{\mu\nu}(X)\p_\pm X^\mu\p_\pm X^\nu=0,
\end{equation}
written here in worldsheet
lightcone coordinates $\sigma^\pm= (\sigma \pm \tau)/\sqrt{2}$.

We will now expand the action covariantly around a background string
solution $X_B^\mu$ of (\ref{euler}). The standard technique for
this is the Riemann coordinate expansion $X^\mu = X_B^\mu + \xi^\mu$
discussed in detail in the sigma-model context in
\cite{AlvarezGaume:1981hn} and briefly recalled in appendix \ref{riem}.

For the time being, in order to compare the Riemann coordinate expansion
with the Penrose limit, we are only
interested in the lowest non-trivial order of this expansion.
The e.o.m. for the expansion fields $\xi^{\mu}$ (most readily obtained by
expanding and then varying the action) are 
\begin{equation}\label{expeuler}
\nabla_i\nabla^i \xi^\lambda+R^\lambda_{\phantom{\lambda}\mu\rho_1\nu}
\p_i X_B^\mu \p^i X_B^\nu \xi^{\rho_1}=0.
\end{equation}
The corresponding first-order
constraints are calculated by expanding (\ref{constr})
accordingly (\ref{expconstrfull}), and read
\begin{equation}\label{expconstr}
G_{\mu\nu}\nabla_\pm \xi^\mu\p_\pm X_B^\nu=0.
\end{equation}

It is now convenient to introduce a frame $E_A^\mu(X_B)$
along the worldsheet.
The
components tangential to the worldsheet $E_i^\mu$, $i\in\{+,-\}$
or $\{\tau,\sigma\}$, are chosen to be the derivatives along the 
coordinate lines of the conformal gauge coordinate system, viewed as
the stringy generalisation of the geodesic affine parameter, i.e.
\begin{equation}
\label{frame1}
E_i=\p_i,\quad E_i^\mu=\p_i X_B^\mu,
\end{equation}
completeted by an orthonormal frame $E_a^\mu$, $a\in\{2,...,D-1\}$ (determined 
up to transverse orthogonal frame rotations), such that
\begin{equation}
\label{frame2}
G_{\mu\nu}E_i^\mu E_j^\nu=g_{ij},\quad G_{\mu\nu}E_i^\mu E_a^\nu=0,
\quad G_{\mu\nu}E_a^\mu E_b^\nu=\delta_{ab}.
\end{equation}
Thus $g_{ij}$ is the induced metric on the classical 
worldsheet background (constrained to be
conformally flat by the conformal gauge condition). 
The string e.o.m  (\ref{euler}) can now simply be written as
$\nabla^i E_i=0$,
replacing the auto-parallelity condition $\nabla_\tau E_\tau=0$ of a geodesic. 
They can be supplemented by the integrability conditions
$\epsilon^{ij}\nabla_iE_j=0$,
which are due to the fact that the $E_i$ are coordinate vectors. 
In terms of the worldsheet lightcone coordinates $\sigma^\pm$, 
these two equations can then be written in the condensed and useful form 
\begin{equation}\label{easy}
\nabla_\pm E_\mp=0.
\end{equation}
After decomposition of the expansion fields into their tangential and
transverse components,
\begin{equation}
\xi^\mu=\xi^A E_A^\mu = \xi^i E_i^\mu +\xi^a E_a^\mu,
\end{equation}
one can reformulate the action, e.o.m.\ and the constraints in frame
component form. Using (\ref{frame2}) and (\ref{easy}), we find for the latter
\begin{equation}\label{expconst}
g_{+-} \partial_\pm \xi^\mp-G_{\mu\nu}E_a^\mu \nabla_\pm E_\pm^\nu \xi^a =0.
\end{equation}
These constraints can be solved for the (tangential, longitudinal) 
lightcone components $\xi^\pm$, up to the residual gauge freedom
$\xi^\pm \ra \xi^\pm + f^\pm(\sigma^\pm)$.
Therefore
their e.o.m. must be satisfied identically by virtue of the constraints. 
Indeed,
after a lengthy calculation we find that the tangential components of
(\ref{expeuler}) are just the derivatives of (\ref{expconst}), i.e.
\begin{equation}
\partial_\mp(g_{+-}\partial_\pm \xi^\mp-G_{\mu\nu} E_a^\mu 
\nabla_\pm E_\pm^\nu \xi^a)=0.
\end{equation}
Furthermore, since the tangential components $\xi^\pm$ appear 
in the transverse components of the e.o.m.\ (\ref{expeuler})
\begin{multline}
\label{treom}
\partial_+\partial_-\xi^a 
+ G_{\mu\nu} \left( E^{a\mu} \nabla_+E_+^\nu \partial_-\xi^+   
+  E^{a\mu} \nabla_+E_b^\nu \partial_-\xi^b 
+   E^{a\mu}  \nabla_-E_-^\nu \partial_+ \xi^-  
+ E^{a\mu} \nabla_-E_b^\nu \partial_+\xi^b\right) \\
+ \frac{1}{2} G_{\mu\nu} \left( E^{a\mu}  \nabla_+\nabla_-E_b^\nu \xi^b 
+ E^{a\mu}  \nabla_-\nabla_+E_b^\nu \xi^b \right)
+  \frac{1}{2} R^a_{\phantom{a}+b-}\xi^b
+ \frac{1}{2} R^a_{\phantom{a}-b+}\xi^b=0
\end{multline}
only via their derivatives $\p_\pm\xi^\mp$, 
we can use the constraints (\ref{expconst}) to completely eliminate
them. 
One then 
finds the purely transverse e.o.m.
\begin{multline}\label{final2}
\left(\frac{1}{2} \del^i\del_i \delta_b^a\phantom{\frac{1}{2}}
+G_{\mu\nu}E^{a\mu} \nabla^i E_b^\nu \partial_i\right.
+G_{\mu\nu}\nabla^i E^{a\mu}\nabla_i E_b^\nu-(G_{\mu\nu} 
\nabla^i E^{a\mu} E_c^\nu)(G_{\lambda\kappa} 
E^{c\lambda} \nabla_i E_b^\kappa)\\
\left.+ \frac{1}{2} G_{\mu\nu} E^{a\mu} \nabla^i\nabla_i E_b^\nu + 
\frac{1}{2} R^{ai}_{\phantom{ai}bi}\right)\xi^b=0.
\end{multline}
Thus we have shown that, to first order in an expansion around a
classical string configuration $X_B$ in an arbitrary curved background,
the tangential fluctuations can be explicitly solved for and eliminated
from the e.o.m. for the true dynamical transverse degrees of freedom by
virtue of the constraints \eqref{expconst}. 

We conclude this section with two comments on these observations:

\begin{enumerate}

\item 
First of all, the fact that the tangential components $\xi^\pm$ can,
in principle, be eliminated to first order is of course related to 
the underlying
worldsheet diffeomorphism invariance. The crucial point here is that
\eqref{expconst} shows how they can explicitly, and thus in practice,
be eliminated in the already partially gauge fixed (conformal gauge)
theory. This should be contrasted with the world-sheet covariant approach,
e.g.\ based on the Nambu-Goto action,
in which the tangential components, identified to first order with
generators of worldsheet diffeomorphisms, can be set to zero (or drop out
of the equations) by virtue of the worldsheet diffeomorphism invariance
(for a geometrically transparent discussion of these issues see e.g.\
\cite{Capovilla:1994bs,Arreaga:2000mr}). However, this is no longer possible
(or true) at higher orders in the expansion,  
which, in contrast to the first order, encode information
beyond mere infinitesimal deviations of nearby strings, and thus are not
(to the same extent) susceptible to worldsheet diffeomorphisms.
Thus if one wants to go
to higher orders (as we will eventually do in section 4), the simplest
way to control the world-sheet diffeomorphisms is to start with a gauge
fixed action and to then simply expand it together with the constraints,
exactly as we have done here to first order.

\item 
Secondly, this is evidently quite reminiscent of the
standard treatment of strings in the lightcone gauge, available for
plane wave (or more general pp-wave or Brinkmann metric) backgrounds
\cite{hs}. We will pursue this analogy in the subsequent
section. To that end it will be useful to rewrite \eqref{final2} in a
manner that makes the underlying geometric structure more manifest,
by introducing the gauge covariant derivative 
w.r.t.\ transverse frame rotations $D_i$
and the extrinsic curvature of the world-sheet $K_{ij}^a$, 
\begin{equation}
D_i \xi^a=\p_i \xi^a +G_{\mu\nu}E^{a\mu} \nabla_i E_b^\nu \xi^b,
\qquad
\label{excur}
K_{ij}^a=G_{\mu\nu} E_i^\mu \nabla_j E^{\nu a}\;\;.
\end{equation}
In terms of these, \eqref{final2} can be written more transparently as
(see e.g.\ \cite{Capovilla:1994bs}) 
\begin{equation}\label{final3}
g^{ij} D_i D_j \xi^a+g^{ij}g^{kl} K_{ik}^a K_{jlb} 
\xi^b+g^{ij}R^{a}_{\phantom{a}jbi}\xi^b=0.
\end{equation}

\end{enumerate}

\section{Transition from strings to null geodesics as background fields}

As mentioned above, the explicit 
elimination of the lightcone degrees of freedom
$\xi^\pm$ from the first order
string expansion by virtue of the constraints
is strikingly reminiscent of the string e.o.m.\ in a
Penrose limit expansion of the metric whose first order is the
plane wave metric
\begin{equation}
\label{pwbc}
ds^2=G_{\mu\nu}dx^\mu dx^\nu
=2dx^+dx^--R_{+a+b}(x^+)x^ax^b dx^+ dx^+ +\delta_{ab} dx^a dx^b.
\end{equation}
Imposing the conformal gauge, the e.o.m.\ for $X^+(\tau,\sigma)$ is
just the free wave equation
\begin{equation}
(\p_\tau^2-\p_\sigma^2)X^+=0,
\label{pwxp}
\end{equation}
and one can fix the residual worldsheet diffeomorphism invariance
by choosing the lightcone gauge
$X^+(\tau,\sigma)=\tau$.
In this gauge, $X^-$ is determined by the constraints
\begin{equation}\label{pwconstr}
\begin{aligned}
\dot{X}^-&-\frac{1}{2} R_{a+b+}X^a X^b+
\frac{1}{2}\delta_{ab} (\dot{X}^a\dot{X}^b+X^{a\prime}X^{b\prime})=0\\
X^{-\prime}&+\delta_{ab}\dot{X}^a X^{b\prime}=0,
\end{aligned}
\end{equation}
and its e.o.m.\ 
\begin{equation}
\label{pwxm}
(\p_\tau^2-\p_\sigma^2)X^-+2R_{+ab+}\p_\tau X^{a}X^{b}
+\frac{1}{2}\p_+ R_{+ab+} X^a X^b=0
\end{equation}
is then, as in section 2, identically satisfied by virtue of the constraints.
The e.o.m.\ for the remaining transverse variables $X^a$ are simply
\begin{equation}\label{pweuler}
(-\p_\tau^2+\p_\sigma^2) X^a-R^a_{\phantom{a}+b+}(\tau)X^b=0.
\end{equation}
Now these equations are quite similar to the transverse equations of motion
(\ref{final2},\ref{final3}),
the difference between the two being mainly due to the complicated
extrinsic curvature information of the background string $X_B$ encoded
in the second term of (\ref{final3}). 

Thinking of strings as probes of the background geometry, one is tempted
to say that the complicated (extrinsic) geometry of the probe itself
obscures or contaminates the background geometry. This becomes most
obvious in flat space where the first order string expansion equations
about an excited string look much more complicated than the exact string
equations themselves. On top of that, for generic curved backgrounds it
is typically very hard to find even one exact solution $X_B$ of the
non-linear string e.o.m.

It is thus legitimate to ask if there is not a better way to perform
a string expansion, one which rids us of all the (for present purposes largely
superfluous) geometric
information encoded in the extrinsic geometry of the string. Of course
the first guess is to try a simpler background $X_B$, ideally an object
with vanishing extrinsic curvature satisfying the exact string e.o.m.\ and
constraints. All of these conditions are satisfied by choosing
$X_B(\tau,\sigma)=\gamma(\tau)$ to be a null geodesic since
\begin{itemize}
\item for $X_B(\tau,\sigma)=\gamma(\tau)$, the e.o.m. \eqref{euler}
reduce to the geodesic equation;
\item the constraints \eqref{constr} reduce to the condition that this
geodesic be null;
\item the extrinsic curvature (\ref{excur})
of $\gamma(\tau)$ vanishes, since a geodesic
extremises proper time.
\end{itemize}
The validity of the first two statements is obvious. As regards the third
claim, note that in general an extremal submanifold is characterised
by the vanishing of the trace of the extrinsic curvature. For a
one-dimensional object this is equivalent to vanishing of the extrinsic
curvature itself, the condition $K_{\tau\tau}^a=0$ being just another way
of writing the geodesic equation.

As we will see in the following, this choice of background will
remedy all the shortcomings mentioned above and, in the end, lead
to a first order string expansion equation of the form (\ref{pweuler}).

First of all we need to address the issue how to formulate the string
expansion around this somewhat degenerate (because $\sigma$-independent)
string background $X_B(\tau,\sigma)=\gamma(\tau)$. 
%
%
It turns out that simply making the replacement 
$X_B^\mu\ra\gamma^\mu$, 
while retaining the $\tau$ and $\sigma$-dependence of $\xi$,
so that e.g.\ 
\begin{equation}
\label{rule1}
\p_\tau X_B^\mu=\dot \gamma^\mu\qquad
\p_\sigma X_B^\mu=\p_\sigma \gamma^\mu=0 \qquad  
\nabla_\sigma \xi^\mu=\p_\sigma \xi^\mu,
\end{equation}
yields valid
expansions of the action, constraints and the e.o.m. 
Therefore we get from
(\ref{expeuler}) the e.o.m.
\begin{equation}\label{expeulerg}
(-\nabla_\tau^2+\p_\sigma^2) \xi^\lambda
-R^\lambda_{\phantom{\lambda}\mu\rho_1\nu}\dot\gamma^\mu \dot\gamma^\nu 
\xi^{\rho_1}=0
\end{equation}
while the constraints (\ref{expconstr}) reduce to
\begin{equation}
\label{expconstrg}
G_{\mu\nu}\nabla_\tau \xi^\mu\dot\gamma^\nu=0 \qquad\qquad
G_{\mu\nu} \p_\sigma \xi^\mu\dot\gamma^\nu =0.
\end{equation}
Using the geodesic equation of motion, these constraints can be 
integrated to
$G_{\mu\nu}\xi^\mu\dot\gamma^\nu=c$
with some constant $c$. We will now show that this constant can be
set to zero. Assume a general solution $\xi(\tau,\sigma)$ of the
e.o.m. (\ref{expeulerg}) and the constraints (\ref{expconstrg}), and
consider the shifted expansion vector
$\tilde{\xi}(\tau,\sigma)=\xi(\tau,\sigma)-c \xi_0(\tau)$,
where $\xi_0(\tau)$  satisfies the ordinary geodesic deviation 
equation with respect to
$\gamma$, and 
is normalised according to
$G_{\mu\nu}\xi_0^\mu\dot\gamma^\nu=1$.
Then $\tilde{\xi}^\mu$ still satisfies the e.o.m.\ (\ref{expeulerg}), but the
constraint is
\begin{equation}\label{intconstr}
G_{\mu\nu}\tilde{\xi}^\mu\dot\gamma^\nu=0.
\end{equation}
In the following we consider two solutions of the first order string
expansion to be equivalent if they differ only by a solution of the mere
geodesic deviation equation, corresponding essentially just to a rigid
displacement of the background geodesic, and consistently set $c=0$.  

Further simplifications arise after introduction of a parallel transported
quasi-orthonormal
frame $E^A_{\mu}$ (with $E^\mu_+=\dot \gamma^\mu$) 
along the null geodesic $\gamma$, as in \eqref{ponf},
since one then has, expanding $\xi^\mu=\xi^AE_A^\mu$ in this basis, 
$\nabla_\tau\xi^\mu=(\p_\tau \xi^A)E_A^\mu$,
so that all covariant derivatives can be replaced by partial
derivatives acting on the frame components.
Hence in frame components the e.o.m. (\ref{expeulerg}) are
simply
\begin{equation}
(\p_\tau^2-\p_\sigma^2)\xi^A+R^A_{\phantom{A}+B+}\xi^B=0,
\end{equation}
while the choice $c=0$ (\ref{intconstr}) is tantamount to
$\xi^-(\tau,\sigma)=0$.
This condition is strictly analogous to the standard condition 
one imposes in the construction
of the transverse geodesic deviation matrix \cite{HE} 
($Z^-=0$ in the notation of \cite[section 2.1]{bbop2}).
Thus, for the individual frame components one finds
\begin{align}
(\p_\tau^2-\p_\sigma^2)\xi^+&=-R^+_{\phantom{+}+B+}\xi^B=-
R^+_{\phantom{+}+-+}\xi^--R^+_{\phantom{+}+a+}\xi^a=-R^+_{\phantom{+}+a+}\xi^a\\
(\p_\tau^2-\p_\sigma^2)\xi^-&=-R^-_{\phantom{-}+B+}\xi^B\equiv 0\\
(\p_\tau^2-\p_\sigma^2)\xi^a&=-R^a_{\phantom{a}+B+}\xi^B
=-R^a_{\phantom{a}+-+}\xi^- -R^a_{\phantom{a}+b+}\xi^b
=-R^a_{\phantom{a}+b+}\xi^b\label{transverse}.
\end{align}
In particular, the transverse equations (\ref{transverse}) are now 
identical to the exact 
transverse string equations (\ref{pweuler}) in a plane wave
background. As regards the equation for $\xi^-$, on the other hand,
comparison with the exact equation (\ref{pwxm})
shows that $\xi^-=0$ is only a solution to the e.o.m.\ to lowest order in the
Riemann expansion - consistent with the fact that in the scaling
\eqref{scaling}
leading to the Penrose plane wave
limit $X^-$ is treated as higher order relative to the $X^a$.

We conclude that the exact transverse string equations in the
first order Penrose-Fermi expansion of the metric $G_{\mu\nu}$ around
$\gamma$, i.e.\ in the Penrose limit plane wave metric associated to
$G_{\mu\nu}$ and $\gamma$, are equivalent to the transverse first-order
string equations obtained by expanding the string embedding fuctions
around a null geodesic $\gamma$ in the original background $G_{\mu\nu}$.


\section{The correspondence to all orders}

To what degree and for which metric/geodesic backgrounds can we expect
the correspondence between the string expansion and
the Penrose-Fermi expansion, which we established above
to first order, to be valid at higher orders? To answer this question
it is worthwhile to take a step back and compare the geometric set-up
in both cases. Although the underlying interpretation is that of an
expansion of the embedding variables on the one hand, and of the metric
on the other, in the end it all reduces to a different prescription
for how to describe the locus of nearby strings in terms of geodesic
distance. This is mirrored by the different adapted coordinate systems
used, i.e.\ Riemann vs.\ Fermi coordinates.

The Riemann coordinates $\xi^+$, $\xi^-$ and  $\xi^a$, used as the
embedding variables in the string expansion, describe the instantaneous
distance to a lightlike particle $\gamma(\tau)$. The somewhat awkward
feature of this coordinate system (in the present context) is that,
as this particle moves
along $\gamma$, these coordinates changes (differentiably) with the affine
parameter, i.e.\ with time.

The Penrose-Fermi expansion, on the other hand, is based on Fermi
coordinates $x^+$, $x^-$ and $x^a$ adapted to the null geodesic $\gamma$
\cite{bfwf}. In Fermi coordinates, one measures distance w.r.t.\ the
null geodesic as a one-dimensional object. To this end space-time
is foliated into transverse hypersurfaces which are parametrised by
the affine parameter, promoted to the Fermi coordinate $x^+=\tau$, and
covered with $D-1$ dimensional, time-independent Riemann coordinates $x^-$
and $x^a$ around the intersection point of geodesic and hypersurface.

At a given but fixed time $\tau=\tau_0$, the position of the string is
described by
\begin{equation}
X^\mu(\tau_0,\sigma)=\gamma^\mu(\tau_0)+\Delta X^\mu(\xi((\tau_0,\sigma)),
\end{equation}
and generically $\xi^{\mu}(\tau_0,\sigma)$ will not lie in the
corresponding transverse hypersurface, because the string is not comoving
with the null geodesic. In that case, the first construction (Riemann
coordinates), in which one simply has 
$X^\mu(\tau_0,\sigma)=\gamma^\mu(\tau_0)+\xi^\mu(\tau_0,\sigma)$
\eqref{ncev},
is more convenient and efficient
than the Fermi construction, as it
accounts for the free movement of the string in space-time.

However, this discussion also shows that both approaches should
agree completely if the string is actually confined to comove with
the null geodesic. To make this more precise, note that comovement
in terms of Fermi coordinates is equivalent to
\begin{equation}\label{lcg2}
X^+(\tau,\sigma)=\tau,
\end{equation}
i.e.\ precisely the lightcone gauge condition, 
whereas in the Riemann string expansion it simply means 
\begin{equation}\label{lcg}
\xi^+(\tau,\sigma)=0.
\end{equation}
Now, by construction 
the transverse Fermi coordinates $(x^{\bar{a}})=(x^-,x^a)$
are equal to the remaining transverse Riemann coordinates (\ref{xtrans}),
\begin{equation}\label{hilf2}
x^{\bar{a}}=\xi^{\bar{a}}_{\gamma(\tau)}.
\end{equation}
Thus for comoving strings (lightcone gauge), the two prescriptions to
measure the locus of the string, namely transverse distance from
the geodesic, indeed agree.
In that special case it is enlighting to recalculate the manifest
covariant form of the string expansion using Fermi and not Riemann
coordinates. As we will show, this significantly simplifies the
identification of the tensorial structures at intermediate steps of the
calculation, and demonstrates that Fermi coordinates are the
ideal reference system to describe the perturbative string expansion
in the lightcone gauge.

To see this, recall first that in Riemann coordinates one has the simple
relationship 
$X^{\mu}(\gamma,\xi)=\gamma^\mu + \xi^\mu$
for the embedding variable, while the expression for its $\tau$-derivative
is more complicated (esentially because Riemann coordinates are anchored at
a fixed basepoint and thus change as one moves along $\gamma$) and given 
by the infinite series (\ref{geodesicd},\ref{geodesicdr}).

In Fermi coordinates, on the other hand, the initial expression for
the expansion of $X^A(\gamma,\xi)$ is somewhat 
more complicated, being given by the infinite series (\ref{fermi1}),
but since this expression holds along the entire null geodesic, no new terms
are generated when taking the $\tau$-derivative \eqref{fermi2}.
The simple (but crucial) observation is now that, 
upon using (\ref{lcg}), this expansion 
\eqref{fermi1} collapses to the simple result
\begin{equation}\label{hilf}
X^A(\gamma,\xi)=\delta^A_+ \tau +\delta^A_{\bar{a}}\xi^{\bar{a}},
\end{equation}
in accordance with (\ref{lcg2}) and (\ref{hilf2}) and the statement
that on the transverse hypersurface $\xi^+=0$ through the event
$\gamma(\tau)$ Fermi coordinates are identical to Riemann coordinates
around $\gamma(\tau)$. Moreover, as a Fermi expression, (\ref{hilf}) is valid
not only at a certain time $\tau$ but all along $\gamma$. Therefore its
time derivative does not include new terms and one simply has
\begin{equation}
\label{txafc}
\p_\tau X^A(\gamma,\xi)=\delta^A_+ +\delta^A_{\bar{a}}\p_\tau \xi^{\bar{a}}.
\end{equation}
as well as (evidently)
\begin{equation}
\label{sxafc}
\p_\sigma X^A(\gamma,\xi)=\delta^A_{\bar{a}}\p_\sigma \xi^{\bar{a}}.
\end{equation}

Thus, provided that one imposes the lightcone gauge one can
simultaneously use the attractive features of Riemann and Fermi
coordinates, i.e.\ one can eat one's cake and have it too, and the
covariant
expansions of $X^A(\tau,\sigma)$ (\ref{hilf}) and its derivatives
(\ref{txafc},\ref{sxafc}) become as simple as they could possibly be.

Moreover, by virtue of the identification (\ref{hilf2}), the tranverse
$\xi^+=0$ Riemann coordinate expansion \eqref{metric} of the metric in terms of
$\xi^{\bar{a}}$ is equivalent to the expansion (\ref{fermimetric})
of the metric in Fermi coordinates. 

Note that, in order to arrive at this conclusion, we only needed to impose
the space-time diffeomorphism gauge condition that the metric be written
in Fermi coordinates as well as the worldsheet diffeomorphism lightcone
gauge condition $X^+=\tau$. This is always possible.

Putting everything together, we conclude that in this combined 
lightcone (worldsheet) and Fermi (space-time) gauge, 
the expansion of the string e.o.m.\ around the null geodesic $\gamma$
becomes identical, to all orders, actually term by term, to the lightcone
gauged string theory e.o.m.\ in the Fermi coordinate expansion of the
metric. Since the expansions agree term by term, this conclusion is
valid both for the ordinary Fermi expansion \eqref{fermimetric} 
as well as for the Penrose-Fermi expansion \eqref{fermipmetric}
of the metric (whose lowest order term is the Penrose limit plane wave)
because the latter is in essence just a reordering of the former.

Frequently, the lightcone gauge is imposed in conjunction with
the conformal gauge, and this imposes strong constraints on the 
background
geometry which lead to the usual simplifications in the subsequent
canonical quantisation. It is well known that the metrics for
which the lightcone gauge can be imposed in addition to the conformal 
gauge are metrics of the Brinkmann form \eqref{brinkmann}
admitting a parallel null vectorfield $\del_v$ \cite{hs}.
Thus, if we insist on the conformal gauge (depending on the form
of the metric, there may also be other suitable gauge choices leading
to a tractable canonical formalism, see e.g.\ \cite{rudd}), we need
to understand for which Brinkmann metrics we can introduce Fermi
cooordinates compatible with the above Brinkmann form. In appendix
\ref{fermbrink} we establish the optimal result along these lines,
namely that 
demanding the Fermi gauge, associated with any one of a spacetime filling
congruence of null geodesics, imposes no further restrictions on
the metric beyond those required by the lightcone and conformal
gauge alone.

\section{Example: Riemann expansion of the plane wave string equations}

To illustrate the above argument regarding the equivalence of the Riemann
and Penrose-Fermi expansions, as a simple example we reconsider the
plane wave in Brinkmann coordinates \eqref{pwbc}. These Brinkmann
coordinates are Fermi coordinates for the central null geodesic
$x^+=\tau,x^{\bar{a}}=0$, and the exact string e.o.m.\ and constraints,
given in (\ref{pwxp})-(\ref{pweuler}), 
are at most quadratic in the transverse fields $X^{\bar{a}}$. Their Riemann 
coordinate expansion, on the other hand, is \textit{a priori} given by an
infinite series.
Thus our claim that these two expansions are (term by term)
equivalent may at first appear to be puzzling. 

To see what is going on, let us take a closer look at the second order
Riemann coordinate string expansion of the e.o.m. (\ref{expeulerfull})
around the null geodesic. Using the rules (\ref{rule1}), one finds
\begin{multline}
(-\nabla_\tau^2+\p_\sigma^2) \xi^\lambda-
R^\lambda_{\phantom{\lambda}\mu\rho_1\nu}\dot\gamma^\mu \dot\gamma^\nu 
\xi^{\rho_1}\\
-2R^\lambda_{\phantom{\lambda}\rho_1\rho_2\mu} 
\dot\gamma^\mu\nabla_\tau \xi^{\rho_1}\xi^{\rho_2}
-\frac{1}{2}\left[\nabla_{\rho_1}R^\lambda_{\phantom{\lambda}\mu\rho_2\nu}
+\nabla_\mu R^\lambda_{\phantom{\lambda}\rho_1\rho_2\nu}\right]
\dot\gamma^\mu \dot\gamma^\nu \xi^{\rho_1} \xi^{\rho_2}+ \mathcal{O}((\xi)^3)=0.
\end{multline}
Evaluating these in frame
components, using the fact that for a plane wave the only nonvanishing
component of the Riemann tensor is $R_{a+b+}(\tau)$, 
and after imposing the lightcone gauge $\xi^+=0$
(\ref{lcg}), one finds the e.o.m.
\begin{equation}
\label{ddxi}
\begin{aligned}
(\p_\tau^2-\p_\sigma^2)(\xi^+=0)+ \mathcal{O}((\xi)^3)=0\\
(\p_\tau^2-\p_\sigma^2)\xi^-+2R_{+ab+}\p_\tau \xi^{a}\xi^{b}
+\frac{1}{2}\p_+ R_{+ab+} \xi^a \xi^b+ \mathcal{O}((\xi)^3)=0\\
(\p_\tau^2-\p_\sigma^2)\xi^a+R^a_{\phantom{a}+b+}\xi^b+ \mathcal{O}((\xi)^3)=0
\end{aligned}
\end{equation}
and similarly the constraints 
\begin{equation}
\begin{aligned}
\dot{\xi}^--\frac{1}{2} R_{a+b+}\xi^a \xi^b+\frac{1}{2}\delta_{ab} 
(\dot{\xi}^a\dot{\xi}^b+\xi^{a\prime}\xi^{b\prime})+ \mathcal{O}((\xi)^3)&=0\\
\xi^{-\prime}+\delta_{ab}\dot{\xi}^a \xi^{b\prime}+ \mathcal{O}((\xi)^3)&=0.
\end{aligned}
\end{equation}
These equations are identical to the standard e.o.m.\ and constraints
in Brinkmann/Fermi coordinates provided that all the higher order
$\mathcal{O}((\xi)^{n\geq 3})$ terms in the Riemann expansion vanish.
Thus the result of section 4 tells us that these terms have to be 
identically zero.

As a check on this geometric reasoning, in this case one can also
establish the absence of these higher order terms in the Riemann
coordinate expansion directly, by using some elementary combinatorial
considerations similar to the kinds of arguments that are used to show
\cite{tseytlin} that plane wave (or pp-wave) backgrounds are exact
solutions of string theory.
Namely, as $\xi^+=0$, there are at most two contravariant $+$ indices,
stemming from $\dot\gamma=E^+$. An initial $R_{+a+b}$ contributes two
covariant indices. Each additional power of the Riemann tensor adds
another two covariant $+$ indices (since contractions are only possible
over transverse indices), and each covariant derivative adds at least
one, namely the $+$-derivative (the others add two as can be seen by
direct inspection of the Christoffel symbols). One covariant $+$ might
be a free contravariant $-$ index (in the e.o.m.\ for $\xi^-$).  Thus,
denoting by $r$ the number of Riemannn tensors and by $d$ the number of
derivatives, we find the condition
\be 
2r + d -1 \leq 2.
\ee
This implies that only terms with $r\leq 1$ and $d\le 1$ can
contribute, thus providing an alternative argument to the effect
that the higher order terms in the expansion \eqref{ddxi} are zero.

\section{Outlook}

For practical applications, the key consequence of our work is the
observation that in the combined Fermi/lightcone gauge, the naive
expansion of the string coordinates (\ref{hilf}) and their derivatives
(\ref{txafc},\ref{sxafc}) is manifestly covariant. This should provide
additional insight into, and significant simplification of, calculations 
performed e.g.\ in the AdS/CFT context (e.g.\ by extending the Fermi
expansion of $AdS_5 \times S^5$ \cite{bfwf} to a string theory expansion).

Applications of this procedure are, however, not limited to the Penrose
limit AdS/CFT context.  For example, it was noted in \cite{thgs} that
the Penrose-Fermi expansion developed in \cite{bfwf}, with $\gamma$
interpreted as a photon trajectory, provides the ideal setting
for performing certain QED calculations (like vacuum polarisation)
in a curved background. It was also remarked there that it would
be interesting to perform analogous calculations in string theory.
We expect the formalism that we have developed in this paper, a stringy
generalisation of \cite{bfwf}, to be useful for that purpose.

The results obtained here also shed light on the propagation of strings in
curved (and singular) backgrounds. For example, some of the observations
in \cite{tolley} regarding the string propagation through a big crunch /
big bang singularity (namely that in the neighbourhood of such a
cosmological singularity the string equations reduce to those in a plane
wave) can be understood as a particular manifestation of the more general
phenomenon that we have described here, since the plane wave in question
is precisely the kind of singular homogeneous plane wave \cite{mmhom}
that was shown in \cite{bbop1,bbop2} to arise generically as the Penrose
limit of a space-time singuarity.

\subsection*{Acknowledgements}

This work has been supported by the Swiss National Science Foundation
and by the EU under contract MRTN-CT-2004-005104.

\appendix

\section{Taylor Expansion in Riemann and Fermi coordinates}\label{riem}

\subsection{Riemann expansion}\label{riem1}

The covariant expansion of a general space-time tensor using Riemann
coordinates is discussed in detail in \cite{AlvarezGaume:1981hn}. Here we
can restrict ourselfes to the embedding variables and the metric. First
note that a coordinate difference $\Delta x^\mu=x^\mu-x_B^\mu$ of
(nearby) points on the curved space-time manifold is an
object whose transformation under space-time diffeomorphisms is not well
defined. Thus a naive Taylor expansion in $\Delta x^\mu$ is bound to
produce correct but nevertheless non-covariant equations. To circumvent
this difficulty one can reparametrise $\Delta x^\mu(\xi)$ by a vector
$\xi$ sitting at $x_B$ by means of the exponential map
\begin{equation}\label{exp}
x^\mu(x_B,\xi)=x_B^\mu+\Delta x^\mu(\xi)=(\Exp_{x_B}(\xi))^\mu.
\end{equation}
As $\xi^\mu$ transforms as a vector, the ordinary
Taylor expansion of the metric in terms of $\xi^\mu$, 
\begin{equation}\label{covten}
G_{\mu\nu}(x_B+\Delta x(\xi))=
\sum_{n=0}^\infty\frac{1}{n!}\frac{\p}{\p\xi^{\rho_1}}
\cdots \frac{\p}{\p\xi^{\rho_n}}G_{\mu\nu}(x_B)\xi^{\rho_1}\cdots\xi^{\rho_n},
\end{equation}
has to be covariant, i.e. the coefficients can be re-expressed in terms
of the curvature tensor and its covariant derivatives.
Note, however, that in a general coordinate system the definition
via the exponential map leads to a rather complicated dependence of
$\Delta x(\xi)$ on $\xi$, namely
\begin{equation}\label{geodesic}
x^\mu(x_B,\xi)=x^\mu_B+\Delta x^\mu(\xi)
=x^\mu_B+\xi^\mu-\sum_{n=2}^\infty\frac{1}{n!}\Gamma_{\rho_1\cdots\rho_n}^\mu 
\xi^{\rho_1}\cdot\xi^{\rho_n},
\end{equation}
where 
$\Gamma_{\rho_1\cdots\rho_n}^\mu=
\nabla_{\rho_1}\ldots\nabla_{\rho_{n-2}}\Gamma_{\rho_{n-1}\rho_{n}}^\mu$
and $\nabla_{\rho}$ means covariant differentiation w.r.t.\ lower indices
only. We see that in order to evaluate (\ref{covten}) one would also have to
expand the coordinate functions $x^\mu$ themselves.

The solution to this problem is to promote $x_B$ to be the origin of a new
coordinate system $\xi^\mu$ in which geodesics emanating from $x_B$ are
straight lines. 
In these Riemann coordinates by definition one has
$\Delta x^\mu=\xi^\mu$ or, equivalently,
\begin{equation}\label{ncev}
x^\mu(x_B,\xi)=x_B^\mu+\xi^\mu,
\end{equation}
making them the natural choice of coordinate system to evaluate
(\ref{covten}). Comparison of (\ref{geodesic}) and (\ref{ncev}) shows
that the symmetrised covariant derivatives of the Christoffel symbols
vanish in Riemann coordinates, 
$\Gamma_{(\rho_1\cdots\rho_n)}^\mu=0$.
From this relation one can iteratively express the 
partially symmetrised derivatives of the 
Christoffel symbols to arbitrary order in terms of the 
Riemann tensor, 
and then use these expressions to manifestly covariantise the expansion
(\ref{covten}), leading to
\begin{equation}\label{metric}
G_{\mu\nu}(x_B+\xi)=G_{\mu\nu}(x_B)-
\frac{1}{3}R_{\mu\rho_1\nu\rho_2}\xi^{\rho_1}\xi^{\rho_2}
-\frac{1}{3!}\nabla_{\rho_1}R_{\mu\rho_1\nu\rho_2} 
\xi^{\rho_1}\xi^{\rho_2}\xi^{\rho_3}+\mathcal{O}((\xi)^4).
\end{equation}
As a tensorial equation, this is now valid in any coordinate system. 

We also need to evaluate the derivative of the embedding variables
$X^{\mu}$, i.e.\ of the  expansion (\ref{geodesic}). Here it is
important to note that, while the symmetrised derivatives of the
Christoffel symbols vanish in Riemann coordinates, this is not true
for their ordinary derivatives. Therefore the derivative of
(\ref{geodesic}) w.r.t. some parameter $\tau$, e.g.\ along a curve
in space-time, leads to an infinite series in Riemann coordinates,
\begin{equation}\label{geodesicd}
\p_\tau X^\mu(X_B,\xi)=\p_\tau(X_B^\mu+\Delta X^\mu(\xi))
=\p_\tau X_B^\mu+\p_\tau \xi^\mu-\sum_{n=2}^\infty\frac{1}{n!}
\left(\p_\nu\Gamma_{\rho_1\cdots\rho_n}^\mu\right) 
\xi^{\rho_1}\cdots\xi^{\rho_n} \p_\tau X_B^\nu.
\end{equation}
In manifestly covariant form this reads
\begin{multline}\label{geodesicdr}
\p_\tau X^\mu(X_B,\xi)=\p_\tau (X_B^\mu+\xi^\mu)
=\p_\tau X_B^\mu+\nabla_\tau\xi^\mu\\ +\left[-\frac{1}{3} 
R^\mu_{\phantom{\mu}\rho_1\nu\rho_2}\xi^{\rho_1}\xi^{\rho_2}
+\frac{1}{12} \nabla_{\rho_1}
R^\mu_{\phantom{\mu}\rho_2\rho_3\nu}\xi^{\rho_1}\xi^{\rho_2}\xi^{\rho_3}
\right]\p_\tau X_B^\nu+\mathcal{O}((\xi)^4).
\end{multline}
Putting everything together, 
we can now write down the expansion of the string e.o.m.\
\eqref{euler},
\begin{multline}\label{expeulerfull}
\nabla_i\nabla^i \xi^\lambda+R^\lambda_{\phantom{\lambda}\mu\rho_1\nu}
\p_i X_B^\mu \p^i X_B^\nu \xi^\rho_1
+2R^\lambda_{\phantom{\lambda}\rho_1\rho_2\mu} 
\p_i X_B^\mu\nabla^i \xi^{\rho_1}\xi^{\rho_2}\\
+\frac{1}{2}\left[\nabla_{\rho_1}
R^\lambda_{\phantom{\lambda}\mu\rho_2\nu}
+\nabla_\mu R^\lambda_{\phantom{\lambda}\rho_1\rho_2\nu}\right]\p_i 
X_B^\mu \p^i X_B^\nu \xi^{\rho_1} \xi^{\rho_2}+\mathcal{O}((\xi)^3)=0.
\end{multline}
and of the constraints (\ref{constr}), 
\begin{equation}
\begin{aligned}\label{expconstrfull}
G_{\mu\nu}(2\nabla_\tau \xi^\mu\p_\tau X_B^\nu 
+2\nabla_\sigma \xi^\mu\p_\sigma X_B^\nu +\nabla_\tau \xi^\mu 
\nabla_\tau \xi^\nu+&\nabla_\sigma \xi^\mu \nabla_\sigma \xi^\nu)\\
-R_{\mu\rho_1\nu\rho_2}\xi^{\rho_1}\xi^{\rho_2} 
(\p_\tau X_B^\mu\p_\tau X_B^\nu+&\p_\sigma X_B^\mu\p_\sigma X_B^\nu)
+\mathcal{O}((\xi)^3))=0\\
 G_{\mu\nu}(\nabla_\tau \xi^\mu\p_\sigma X_B^\nu 
+\nabla_\sigma \xi^\mu\p_\tau X_B^\nu 
+\nabla_\tau \xi^\mu \nabla_\sigma \xi^\nu)
-R_{\mu\rho_1\nu\rho_2}&\xi^{\rho_1}\xi^{\rho_2}\p_\tau X_B^\mu
\p_\sigma X_B^\nu +\mathcal{O}((\xi)^3))=0.
\end{aligned}
\end{equation}

\subsection{Fermi expansion}\label{Fermi}

Riemann coordinates are most suitable to evaluate covariant Taylor
expansions around a point in space-time. However, if one wishes to
expand only transversally to a given geodesic $\gamma$, i.e.\ a
one-dimensional object, Fermi coordinates are the most adequate
tool.  In the following we will restrict the discussion to the case
of null Fermi coordinates, i.e.\ with $\gamma$ a null geodesic,
considered in \cite{bfwf} and constructed as follows. First one
introduces a quasi-orthonormal frame $E^A_\mu$,
\begin{equation}
\label{ponf}
ds^2|_\gamma=\eta_{AB}E^AE^B=2E^+E^-+\delta_{ab}E^aE^b
\end{equation}
parallel transported along $\gamma$, 
with $E^\mu_+=\dot \gamma^\mu$. The tranversality condition is then
implemented by
$\xi^\mu_{\gamma(\tau)} E^+_\mu(\gamma(\tau))=\xi^+_{\gamma(\tau)}=0$,
where $\xi^\mu_{\gamma(\tau)}$ is the vector defining the Riemann
coordinate system around the point $\gamma(\tau)$. The r\^ole of $\xi^+$
is now played by the affine parameter of the geodesic $\tau$, promoted
to be the Fermi coordinate
$x^+=\tau$.
The remaining Fermi coordinates are identical to the Riemann coordinates
restricted to the transverse hypersurface, i.e.
\begin{equation}\label{xtrans}
x^{\bar{a}}=\left. E^{\bar{a}}_\mu\xi^\mu_{\gamma(\tau)}\right|_{\xi^+=0}
=\xi^{\bar{a}}_{\gamma(\tau)}.
\end{equation}

In Fermi coordinates, the Christoffel symbols
as well as the symmetrised transverse components of their covariant or
partial derivatives vanish all along $\gamma$, 
\begin{equation}
\label{along3}
\left.\Gamma_{AB}^C\right|_\gamma=
\left.\p_{(\bar{a}_1}\dots\p_{\bar{a}_{n-2}\phantom{)}}
\Gamma_{\phantom{(}\bar{a}_{n-1}\bar{a}_n)}^A \right|_\gamma=0,
\end{equation}
and not only at a certain point, as for Riemann
coordinates. The price we have to pay for this 
is that this is no longer true for the
symmetrised higher derivatives including the geodesic direction (a
lower $+$-index). For example, while one obviously has
$\Gamma^A_{BC,+}=0$ by (\ref{along3}), one calculates e.g.
\begin{equation}\label{gammafermi}
\Gamma^A_{(+B,C)}=R^A_{\phantom{A}(BC)+}.
\end{equation}
Similarly to the Riemann case, the derivatives of the Christoffel symbols
can be used to determine the explicit expansion of the metric in terms of
the components of the Riemann tensor restricted to the geodesic $\gamma$.
To cubic order (for the quartic terms see \cite{bfwf}) one finds
\begin{equation}\label{fermimetric}
\begin{aligned}
ds^2=&\quad 2dx^+dx^- +\delta_{ab} dx^a dx^b \\
     &-R_{+\bar{a}+{\bar{b}}} \ x^{\bar{a}} x^{\bar{b}} (dx^+)^2 
           -\frac{4}{3} R_{+{\bar{b}}\bar{a}{\bar{c}}} 
          x^{\bar{b}} x^{\bar{c}} (dx^+ dx^{\bar{a}}) -\frac{1}{3}
          R_{\bar{a}{\bar{c}}{\bar{b}}{\bar{d}}}  x^{\bar{c}} x^{\bar{d}}
          (dx^{\bar{a}} dx^{\bar{b}})\\
     &-\frac{1}{3} R_{+{\bar{a}}+{\bar{b}};{\bar{c}}} 
          x^{\bar{a}} x^{\bar{b}} x^{\bar{c}} (dx^+)^2   
          -\frac{1}{4} R_{+{\bar{b}}{\bar{a}}{\bar{c}};{\bar{d}}} \ x^{\bar{b}}
          x^{\bar{c}} x^{\bar{d}} (dx^+ dx^{\bar{a}}) 
         -\frac{1}{6} R_{{\bar{a}}{\bar{c}}{\bar{b}}{\bar{d}};{\bar{e}}} 
          x^{\bar{c}} x^{\bar{d}} x^{\bar{e}}(dx^{\bar{a}} dx^{\bar{b}})\\
    & + \mathcal{O}(x^{\bar{a}}x^{\bar{b}}x^{\bar{c}}x^{\bar{d}})
\end{aligned}
\end{equation}
Turning now to the expansion of the coordinates and embedding variables, 
direct insertion of (\ref{gammafermi}) into the expansion
(\ref{geodesic}) leads to
\begin{equation}\label{fermi1}
x^A(\gamma,\xi)=\gamma^A+\Delta x^A(\xi)
=\delta^A_+ \tau +\xi^A-R^A_{\phantom{A}
+\bar{c}+}\xi^+\xi^+\xi^{\bar{c}}
-2 R^A_{\phantom{A}\bar{b}\bar{c}+}\xi^+\xi^{\bar{b}}\xi^{\bar{c}}
+\mathcal{O}((\xi)^3).
\end{equation}
In constrast to the Riemann expansion it contains terms of arbitrary
high order in $\xi^A$ (as long as $\xi^+\ne0$). However this expression
is valid along $\gamma$. Accordingly we find, using
(\ref{along3}), that no new terms appear after differentiation of the 
embedding variables, 
\begin{multline}\label{fermi2}
\p_\tau X^A(\gamma,\xi)=\p_\tau(\gamma^A+\Delta X^A(\xi))\\
=\delta^A_++\p_\tau \xi^A
-\p_\tau(R^A_{\phantom{A}+\bar{c}+} \xi^+\xi^+\xi^{\bar{c}})
-2\p_\tau( R^A_{\phantom{A}\bar{b}\bar{c}+}\xi^+\xi^{\bar{b}}\xi^{\bar{c}})
+\mathcal{O}((\xi)^3)
\end{multline}

\subsection{Penrose-Fermi expansion}\label{FermiP}

In \cite{bfwf} the Fermi expansion of the metric around a
null geodesic was used to define a covariant extension of the Penrose
limit to higher orders, i.e. a Penrose-Fermi expansion. In a nutshell
the prescription is to rescale the Fermi coordinates together with a
conformal transformation of the metric
\begin{equation}
\label{scaling}
(x^+_\lambda,x^-_\lambda,x^a_\lambda)=
(x^+,\lambda^2 x^-,\lambda x^a),\quad ds^2_\lambda=\frac{1}{\lambda^2} ds^2 
\end{equation}
This leads to a reshuffling of the terms in the Fermi expansion whose
zero'th order term in $\lambda$ is the Penrose limit plane wave associated
with the metric $G_{\mu\nu}$ and the null geodesic $\gamma$,
\begin{equation}
\label{fermipmetric}
\begin{aligned}
ds^2 &= \quad 2dx^+dx^- + \delta_{ab}dx^a dx^b -R_{a+b+}x^ax^b (dx^+)^2\\
     &+ \lambda\left[-2 R_{+a+-} \ x^a x^-
(dx^+)^2 -\frac{4}{3} R_{+bac} \ x^b x^c (dx^+ dx^a) 
-\frac{1}{3} R_{+a+b;c} \ x^a x^b x^c (dx^+)^2 \right] +
\mathcal{O}(\lambda^2)
\end{aligned}
\end{equation}

\section{Fermi coordinates compatible with the Brinkmann form}\label{fermbrink}

Here we want to show that there always exist Fermi coordinates
$(x^+,x^-,x^a)$ which are compatible with the general (Brinkmann) form
\begin{equation}\label{brinkmann}
ds^2=2du(dv+A(u,y^k)du+A_i(u,y^k)dy^i)+G_{ij}(u,y^k)dy^idy^j
\end{equation}
of a metric admitting a null parallel (and hence in particular Killing)
vector $\del_v$. 
This means that in
this new coordinate system the metric has the same general form as above,
and moreover has the features that (a) $x^+=\tau$, $x^-=0$, $x^a=0$ is the
basic null geodesic $\gamma$, (b) $\p_+|_\gamma,\p_-|_\gamma,\p_a|_\gamma$
is a quasi-orthonormal parallel frame along $\gamma$, and (c) all the
curves $x^+=c^+$, $x^-=c^- t$, $x^a=c^a t$ with $c^+$, $c^-$, $c^a=const.$
are also geodesics.

In order to identify a suitable null geodesic $\gamma$ (actually, as
we will see, a whole congruence of null geodesics), we first cast the
Brinkmann metric (\ref{brinkmann}) into the Rosen coordinate form
\begin{equation}
ds^2=2dudv+G_{ij}(u,y^k)dy^idy^j,
\end{equation}
which is always possible \cite{tseytlin}. It is now readily checked
that any curve $u=p^v\tau$ , $p^v\ne 0$, with $v,y^i=const.$ is a null
geodesic. Pick one of this congruence, set $p^v=1$, call it $\gamma$,
shift $v$ so that $\gamma$ sits at $(v=0,y^i=y^i_0)$, 
and introduce the corresponding Fermi coordinate $x^+=u=\tau$.

Moreover, $x^+=c^+$ ($p^v=0$) is also a solution to the geodesic
e.o.m.\ and thus the hypersurfaces $x^+=c^+$ can be generated
by transverse geodesics emanating from the intersection point with
$\gamma$. $\p_v$ is parallel and hence, in particular, 
parallel
transported along $\gamma$. Choose $E_+=\dot\gamma$ and $E_-=\p_v$ and
complete it by $E_a=E_a^i\p_i$ to a quasi-orthonomal parallel
frame along $\gamma$. In any one of 
the spacelike codimension 2 surfaces $v,x^+=const.$ 
spanned by the $y^i$, with induced metric $G_{ij}(x^+,y^k)$,
we introduce Riemann normal coordinates $x^a$
around the point $(y^i_0)$ w.r.t.\ the frame $E_a(x^+)$, i.e.\ such
that $\del_a|_\gamma = E_a$. Since $G_{ij}(x^+,y^k)$ is independent of
$v$, this
can be achieved by a $v$-independent, but generically $x^+$-dependent,
coordinate transformation of the
form $x^a=x^a(x^+,y^i)$. Then the metric takes the form
\begin{equation}\label{br3}
ds^2=2dx^+(dv+A(x^+,x^c)dx^++A_a(x^+,x^c) dx^a)+G_{ab}(x^+,x^c)dx^adx^b.
\end{equation}
Note that, while this has the same general form as \eqref{brinkmann}, 
the coordinates are now such that (a)
$x^+=\tau$,  $v=0$, $x^a=0$ is the Fermi null
geodesic $\gamma$, 
and (b) $\p_+,\p_v,\p_a$ is parallel quasi-orthonormal along $\gamma$. 
Furthermore, the geodesic e.o.m.\ for the $x^a$ are satisfied by
$x^a=c^a t$ with $x^+=c^+$, since $A$ and $A_a$ do not contribute for
$\dot{x}^+=0$ and the $x^a$ are spatial Riemann coordinates for $G_{ab}$.

To completely satisfy criterion (c),
we still need to replace $v$ by a coordinate $x^-$ whose geodesic
e.o.m.\ are fulfilled by $x^-=c^- t$, $x^a=c^at$ and $x^+=c^+$ for all
$c^+,c^-,c^a$, and such that $\p_-$ is quasi-orthonormal parallel along
$\gamma$. The only coordinate transformation left to us is a shift
$x^-=v+P(x^+,x^a)$. Note that this shift changes only $A$ and $A_a$
in (\ref{br3}) and therefore does not effect the e.o.m. for $x^a$
if $\dot x^+=0$. Futhermore, if $P$ is at least quadratic in the $x^a$,
the Jacobian of the coordinate transformation is trivial on $\gamma$,
and therefore $\p_+,\p_-,\p_a$ is parallel along $\gamma$ because there
it is identical to the above parallel frame $\p_+,\p_v,\p_a$.

After the shift, the $x^-$ e.o.m.\ is 
\begin{multline}
2\ddot x^-=-\p_a\p_b P(c^+,c^d t)c^ac^b-\frac{d}{dt}(A_a(c^+,c^d t)c^a
+\p_{x^+}G_{ab}(c^+,c^d t)c^ac^b\\
=-\p_a\p_b P(c^+,c^d t)c^ac^b-\p_a(A_b(c^+,c^d t)c^ac^b+\p_{x^+}
G_{ab}(c^+,c^d t)c^ac^b
\end{multline}
where we used $\dot x^+=0$. We want the right side to vanish.
Rescaling $c^a$ by $t$ we get
\begin{equation}
\p_a\p_b P(c^+,c^d)c^ac^b
=-\p_a A_b(c^+,c^d)c^ac^b+\p_{x^+}G_{ab}(c^+,c^d)c^ac^b\equiv
D_{ab}(c^d)c^ac^b.
\end{equation}
Expanding both sides in a Taylor series in the $c^a$, 
comparison of coefficients gives
\begin{equation}
\p_{(a_1}\cdots \p_{a_n)} P(c^+,0)
=\p_{(a_1}\cdots \p_{a_{n-2}} D_{a_{n-1}a_{n})}(c^+,0).
\end{equation}
This can always uniquely be solved for given $D_{ab}$. Finally, as $A_a$
is at least linear in the $x^a$ (the metric restricted to $\gamma$ is flat) 
and $\p_{x^+}G_{ab}$ is at least quadratic (Riemann coordinate metric), 
$P$ is also at least quadratic in the $x^a$, as required.

\rnc{\Large}{\normalsize}

\end{document}